\documentclass{article}

\usepackage{PRIMEarxiv}

\usepackage[utf8]{inputenc} 
\usepackage[T1]{fontenc}    
\usepackage{hyperref}       
\usepackage{url}            
\usepackage{booktabs}       
\usepackage{amsfonts}       
\usepackage{nicefrac}       
\usepackage{microtype}      
\usepackage{lipsum}
\usepackage{fancyhdr}       
\usepackage{graphicx}       

\usepackage{balance} 
\usepackage{multirow}
\usepackage{listings}
\usepackage{subfigure}
\usepackage{subcaption}
\usepackage{mathtools}
\usepackage{makecell}

\usepackage{soul}  
\usepackage{xcolor}
\usepackage[colorinlistoftodos]{todonotes}
\usepackage[export]{adjustbox}
\usepackage{cleveref}

\usepackage{algorithm}
\usepackage[noend]{algorithmic}
\usepackage[numbers]{natbib}

\usepackage{xcolor}
\usepackage{amsmath}
\usepackage{amssymb}
\usepackage{cleveref}
\usepackage{algorithm}
\usepackage[noend]{algorithmic}

\usepackage{multirow}

\usepackage{listings}
\usepackage{xcolor}
\usepackage{authblk}

\newtheorem{definition}{Definition}

\newcommand{\frameworkname}{APThreatHunter}
\newcommand{\threat}{\theta}
\newcommand{\threatsset}{\Theta}
\newcommand{\system}{\digamma}

\newcommand{\domainfile}[1]{\operatorname{Domain}({#1})}

\newcommand{\planscount}{k}
\newcommand{\constructIstate}[1]{\operatorname{ConstructIState}_{#1}}
\newcommand{\constructGstate}[1]{\operatorname{ConstructGState}_{#1}}

\newcommand{\planner}[1]{\operatorname{Planner}_{#1}}

\lstdefinelanguage{PDDL}{
  sensitive=false,
  morecomment=[l]{;},
  alsoletter={:,-,?},           
  moredelim=**[is][\bf]{@}{@},
  morekeywords={
    define,domain,problem,not,and,or,when,forall,exists,either,
    :domain,:requirements,:types,:objects,:constants,
    :predicates,:action,:parameters,:precondition,:effect,
    :fluents,:primary-effect,:side-effect,:init,:goal,
    :strips,:adl,:equality,:typing,:conditional-effects,
    :negative-preconditions,:disjunctive-preconditions,
    :existential-preconditions,:universal-preconditions,:quantified-preconditions,
    :functions,assign,increase,decrease,scale-up,scale-down,
    :metric,minimize,maximize,
    :durative-actions,:duration-inequalities,:continuous-effects,
    :durative-action,:duration,:condition,:constraints
  }
}

\lstnewenvironment{pddllisting}[1][]{
  \lstset{
    language=PDDL,
    basicstyle=\ttfamily\small,   
    columns=fullflexible,         
    keepspaces=true,              
    showstringspaces=false,
    upquote=true,
    captionpos=b,
    frame=single,
    breaklines=true,        
    breakatwhitespace=true, 
    literate={\{}{{\{}}1{\}}{{\}}}1{(}{{(}}1{)}{{)}}1,
    #1
  }
}{}
\lstdefinelanguage{ASP}
{
  sensitive=true,                 
  morecomment=[l]\%,              
  morecomment=[s]{/*}{*/},        
  morestring=[b]",                
  alsoletter={\#,:,-,~,/,@},      
  moredelim=**[is][\bf]{@}{@},    
  morekeywords={
    not,true,false,
    {\#show},{\#const},{\#include},{\#external},{\#program},{\#script},{\#base},
    {\#domain},{\#edge},{\#heuristic},{\#theory},
    {\#sum},{\#count},{\#min},{\#max},{\#maximize},{\#minimize}
  }
}

\lstnewenvironment{asplisting}[1][]
  {
    \lstset{
      language=ASP,
      basicstyle=\ttfamily\footnotesize, 
      breaklines=true,
      columns=fullflexible,
      captionpos=b,
      name=ASP Listing,
      frame=single,
      #1
    }
  }
  {}

\graphicspath{{media/}}     

\title{\frameworkname: An automated planning-based threat hunting framework}

\author[1,3]{Mustafa F. Abdelwahed}
\author[2]{Ahmed Shafee}
\author[1]{Joan Espasa}

\affil[1]{School of Computer Science, University of St Andrews, United Kingdom}
\affil[2]{Department of Engineering and Computer Science, Adams State University, Alamosa, CO, USA}
\affil[3]{EG-CERT, NTRA, Egypt}

\begin{document}
\maketitle

\begin{abstract}
Cyber attacks threaten economic interests, critical infrastructure, and public health and safety. To counter this, entities adopt cyber threat hunting, a proactive approach that involves formulating hypotheses and searching for attack patterns within organisational networks.
Automating cyber threat hunting presents challenges, particularly in generating hypotheses, as it is a manually created and confirmed process, making it time-consuming. 
To address these challenges, we introduce \frameworkname, an automated threat hunting solution that generates hypotheses which eliminates the analyst bias. This is done by using automated planning to generate a set of possible risks based on the system's current state and how such threats can be preformed.
We evaluated \frameworkname{} using real-world android malware samples. Results revealed the practicality of using automated planning for goal hypothesis generation in cyber threat hunting.
\end{abstract}


\section{Introduction and Background}
Cyber attacks pose a substantial threat to economic interests, critical infrastructure, and public health and safety~\cite{GROBYS2022101534,beerman2023review,ghafur2019retrospective}. In response, entities have embraced cyber threat hunting, a proactive approach involving hypothesis formulation and searching for attack patterns—specifically the tactics, techniques, and procedures (TTPs) used by threat actors~\cite{staff2018cyberspace}. However, cyber analysts must manually investigate and reconstruct attacks, making threat hunting time-consuming and subject to analyst bias~\cite{anderwthreat2024}. Threat hypothesis generation remains a crucial challenge facing automation efforts~\cite{ThreatScout}, as reliance on manual creation leads to security fatigue and reduced performance~\cite{nour2023survey}.
In this paper, we introduce \frameworkname{}, an automated threat hunting solution that generates threat hypotheses without human intervention. \frameworkname{} employs logic programming~\cite{lloyd2012foundations} to determine the system's current state based on monitored data points (e.g., system calls) and Automated Planning (AP)~\cite{helmert2008understanding} to identify potential risks. Unlike rule-based detection systems, which require all conditions to be met before triggering an alarm, \frameworkname{} proactively uses a planner to generate potential hypotheses and methods to achieve them, which can then be translated into multiple rules.
As a case study, we implemented \frameworkname{} for Android mobile devices, an underexplored area despite Android accounting for approximately 75\% of the global mobile OS market~\cite{Statcounter} and security companies reporting hundreds of thousands of new mobile malware samples daily~\cite{AVTest}. We utilise \frameworkname{} to detect financial fraud and surveillance threats. For example, consider the dirty COW kernel privilege escalation vulnerability (\texttt{CVE-2016-5195}), which allows attackers to gain root access. \frameworkname{} receives data feeds regarding the APK under examination, reasons on these to infer the device's current state, and employs automated planning to identify potential attack continuations. If a plan is found, this indicates a possible threat, with the plan denoting how the attack could be performed.
%
The two main components of the system are a logic programming solver and an AI planner.

\textbf{Logic Programming} is used to infer an abstraction of the system's current state, based on data feeds.
There are several logic programming languages, such as DataLog~\cite{datalog}, Prolog~\cite{colmerauer1990introduction} and Answer Set Programming (ASP)~\cite{baral2003knowledge}. A logic program defines a problem's description using a set of rules and constraints, while facts represent problem instances. Here we use ASP to describe the system's current state. ASP is a knowledge representation language that supports non-monotonic reasoning capabilities, enabling the removal of assumptions or conclusions. This makes it an ideal choice for common-sense reasoning. The primary construction element in an ASP program is an atom or rule, represented as ${Head}:-{Body}$, indicating that the $Head$ holds if the $Body$ holds. Any ASP program operates as follows, first it grounds variables with their possible values then constructs a set of solutions that satisfies a given set of constraints.

\textbf{Automated Planner} is used to generate potential threats based on the system's current state. 
Following \citet{ghallab2016automated}, a planning task is defined as a tuple $\Xi=\langle S, A, \gamma, I, G\rangle$, where $S$ is a set of states, $A$ is a set of actions, and $\gamma: S \times A \rightarrow S$ is a transition function that associates each state $s\in S$ and action $a\in A$ to the next state $\gamma(s,a)=s^\prime$. $I\in S$ represents the initial state, and $G\in S$ is the goal formula. A solution for $\Xi$ is a plan ($\pi$) defined as a sequence of actions $a_1, a_2, \ldots, a_m$ such that $a_i\in A$ and $\gamma(\gamma(\gamma(I, a_1),\ldots),a_m)=G$. $\Pi_\Xi$ denotes a set of all plans for planning task $\Xi$.

\section{Threat Hunting as Planning}

In this section, we cast the threat hunting problem as a planning problem then propose a generic framework for its operation. Since the framework is independent from the system under examination, we refer to such a system as $\system$ and a domain file (i.e., actions) of $\Xi$ as $\domainfile{\Xi}$. The domain file contains actions were the planner can use to try to perform an attack.
Based on the definition of a planning problem provided by \citeauthor{ghallab2016automated}, we define threat hunting as a planning task as follows:

\begin{definition}[Threat Hunting Task]\label{def:thtask}
    Given a system $\system$ and a planning task's domain $\domainfile{\Xi}$, construct a set of possible threats $\threatsset$ for system $\system$. 
\end{definition}

\begin{figure}
    \centering
    \includegraphics[scale=0.72]{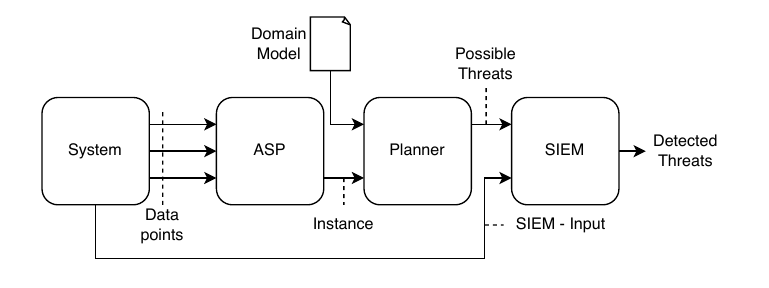}
    \caption{System Pipeline.}
    \label{fig:system-pipeline}
\end{figure}

\Cref{fig:system-pipeline} illustrates the architecture of \frameworkname{}. The ASP component receives data points from the system under examination, reasons on them and then translates the outcomes of this reasoning into a planning problem instance. Subsequently, the planner receives the planning instance along with a domain model that defines the threats and produces a set of possible threats (i.e., $\threatsset$). In general, a Security Information and Event Management (SIEM) solution~\cite{siem-solution} is a cybersecurity system that collects and analyses security data from various sources across an IT environment to detect and respond to threats. To check if any of the generated possible threats in $\threatsset$ exists, the SIEM component translates every plan in $\threatsset$ into a rule that will be triggered if its conditions are met. Such rules are referred to as Indicators of Compromise (IoC). The translation of the rules are left out of scope for this paper.


\Cref{alg:threat-detector} illustrates the operation of \frameworkname, which aims to construct a set of threat hypotheses $\threatsset$. Initially, it starts with an empty set of threat hypotheses $\threatsset$, then constructs the current system's state using $\constructIstate{\Xi}:\{\system\}\rightarrow S$ (Lines:\ref{alg-line:empty-threat-list}-\ref{alg-line:construct-I}). The $\constructIstate{\Xi}$ converts the system's feed into a planning state $I\in S$, while every $\threat$ in $\threatsset$ gets converted into a goal state $G$ using $\constructGstate{\Xi}:\threatsset\rightarrow S$.
After constructing a planning task (i.e., $I$ and $G$), \frameworkname{} invokes a planner to find $k$ plans using $\planner{\Xi}:\mathbb{N}^+\times S \times S \rightarrow \Pi_\Xi$ (Line~\ref{alg-line:find-threat}). The reason behind generating $k$ plans is to deal with the uncertainty of how a threat can be achieved, as there are more than one way to perform the same threat.
For each plan $\pi\in\Pi$, it gets added to the set of possible threats.
%
By reflecting such functions on the system's pipeline, The ASP block implements $\constructIstate{\Xi}$ and $\constructGstate{\Xi}$, while the planner block implements the $\planner{\Xi}$ function. The following section covers a case study showing how \frameworkname{} is implemented for an Android device.

\begin{algorithm}
\caption{$\operatorname{IdenftiyThreats}$}\label{alg:threat-detector}
    \begin{algorithmic}[1]
        \REQUIRE $\system$: System, $\domainfile{\Xi}$: Planning domain,
        $\planscount$: Number of plans per threat
        \ENSURE $\threatsset$ A set of possible threats.
        \STATE  $\threatsset \gets\{\}$ \label{alg-line:empty-threat-list}
        \STATE  $I \gets \constructIstate{\Xi}(\system)$ \label{alg-line:construct-I}
        \STATE  $G \gets \constructGstate{\Xi}(\threat)$ \label{alg-line:construct-G}
        \STATE  $\Pi \gets \planner{\Xi}(\planscount, I, G)$ \label{alg-line:find-threat}
            \FOR{$\pi\in\Pi$} \label{alg-line:confirm-threat-start}
                \STATE $\threatsset\gets\threatsset\cup\{\threat\}$ \label{alg-line:confirm-threat-end}
            \ENDFOR
        \RETURN $\threatsset$
    \end{algorithmic}
\end{algorithm}

\subsection{Threat hunting for Android Devices}

In this work, we focus on two possible threats: financial fraud, and surveillance. Both threats are highly prevalent on Android and cause immediate and measurable harm~\cite{xu_dva,Anglano_survey}. Financial fraud produces direct monetary loss and reputational damage by compromising banking workflows and credentials. Surveillance violates user privacy through misuse of sensors. To implement \frameworkname, we need to define the data feeds, implement $\constructIstate{\Xi}$ and $\constructGstate{\Xi}$. First, we define a state using a set of Boolean predicates shown in \Cref{tbl:asp-inferred-predicates}.  

\begin{table*}[!ht]
\centering
\small
\begin{tabular}{l|l}
Predicate                                    & Description \\ \hline
\texttt{(exploited ?v)}                      & Indicates that a specific vulnerability has been successfully exploited \\
\texttt{(a11y-service-active ?a)}            & Indicates that an application has activated accessibility services \\
\texttt{(notification-accessible ?a)}        & Indicates that an application can access and intercept system notifications\\
\texttt{(clipboard-readable ?a)}             & Indicates that an application can read clipboard content \\
\texttt{(perm-granted ?a ?s)}                & Indicates that an application has been granted permission to access specific sensors \\
\texttt{(cross-sandbox-reads ?a)}            & Indicates that an application has successfully bypassed Android's sandbox security model\\ \hline
\end{tabular}
\caption{Predicates inferred by ASP. Types in predicates are removed for space reasons.}
\label{tbl:asp-inferred-predicates}
\end{table*}

%
%
We convert the APK's system calls, permissions, intends into set of ASP facts. A \emph{system call} is a request from an application to the operating system to perform a privileged operation, such as accessing a file or allocating memory. \emph{Permissions} specify which sensitive resources an  application is allowed to access, such as the camera, location, or contacts. \emph{Intents} are simple messages that applications use to request actions from the  system or other apps, such as opening a screen, starting a service, or sending a broadcast.

Any system call is converted into the following fact: \texttt{invoked(T, N, PID, ARGS, RET)}, where the \texttt{T} holds the time stamp this call was invoked at, \texttt{N} is the name of the system call, \texttt{PID} is the process invoked this system call, \texttt{ARGS} the call's arguments values and \texttt{RET} is the return value of the call. 
To exemplify the translation phase, consider the \texttt{dup} system call. This system call creates a new file descriptor that refers to the same open file as an existing descriptor. We translate this fact, given its C-style declaration \texttt{int dup(int oldfd)}, into \texttt{invoked(T, dup, Pid, OldFd, NewFd, ReturnStatus)}.
This translation is inspired by previous work ~\cite{abdelwahed2023detecting}, where system calls are translated into facts to infer malicious activities for Microsoft Windows binaries. Permissions and intents are translated in a similar way. A permission is encoded as a fact of the form \texttt{has\_permission(Perm)}, indicating that the application either requests (declares) or has access to the sensitive resource \texttt{Perm} (e.g., \texttt{has\_permission(camera)} or \texttt{has\_permission(access\_fine\_location)}). 
Intents are represented as \texttt{intend\_action(Action)} and  \texttt{intend\_category(Category)}. 
For example, \texttt{intend\_action(boot\_completed)}, indicating that the app requests to be notified when the device finishes booting, or \texttt{intend\_category(home)}, indicating that the app declares itself as part of the ``home'' category, such as a launcher). Since all facts in a trace correspond to the same application being analysed, the application identifier is implicit and not included as an argument in these facts.

The operation of $\constructIstate{\Xi}$ is as follows. After the translation is
performed, an ASP model
is used to reason on the generated facts. To clarify this part, we
use the dirty COW \texttt{CVE-2016-5195} as a guiding example. To exploit such a
vulnerability, the attacker first opens a sensitive system file (for example, a
file that controls user accounts or a privileged binary), then uses \texttt{dup}
to duplicate the file descriptor so that the file remains accessible even if the
original descriptor is closed or altered during the exploit. Next, the attacker
calls \texttt{mmap}
to create a writable memory mapping of what is supposed to be a read-only file.
By repeatedly writing to this memory mapping and triggering a race condition in
the kernel's copy-on-write mechanism, the attacker can silently modify the
contents of the protected file. Once the attacker has overwritten this file with
their own data (for instance, adding a new root user or altering a privileged
program), they effectively obtain root privileges on the device. To capture such
behaviour, we create a specific rule for it, and we do the same for all known
vulnerabilities. \Cref{lst:cve-asp} shows several rules to infer such
vulnerability including the one we just covered. If any of system calls in the
rules' bodies are detected with a specific arguments then
\texttt{exploited(cve\_2016\_5195)} and
\texttt{enables\_privilege\_escalation(cve\_2016\_5195)} will hold, then the new
holding facts are rewritten into PDDL. For example, the
\texttt{exploited(cve\_2016\_5195)} gets rewritten into \texttt{(exploited
cve\_2016\_5195)}, same for the
\texttt{enables\_privilege\_escalation(cve\_2016\_5195)}. As for the
$\constructGstate{\Xi}$ operation, it grounds the predicate
\texttt{(threat-possible ?t - threat ?m - mechanism ?a - app)}. This concludes
the ASP block shown in \Cref{fig:system-pipeline}.

Our PDDL defines actions that the planner can use to perform financial fraud or surveillance. A financial fraud threat can be achieved by exploiting accessibility services and system alert windows to intercept credentials, manipulate banking interfaces via overlay attacks, and harvest sensitive financial data through clipboard access and notification interception. A surveillance threat can be achieved by gaining access to hardware sensors such as: camera, microphone, GPS and the phone's screen.
The way a threat is performed is refereed to as a mechanism. In this work, we have two primary mechanisms which are gaining permissions or exploiting vulnerabilities. For generality, the PDDL model is designed to consider several applications running. However, the Android OS is designed to run APKs in a sandbox (i.e., contained environment) making it harder for applications to access each other's memory, thus we assume a single application called \texttt{app} for this case. 

\begin{asplisting}[caption=ASP facts and rules for CVE-2016-5195, label={lst:cve-asp}][!pb]
exploited(cve_2016_5195) :-
    invoked(T1, dup, P, _, _, 0),
    invoked(T2, mmap, P, _, buffer, read_or_write, 0).
...
exploited(cve_2016_5195) :- 
    invoked(T1, read, P, _, buffer, read, 0), 
    invoked(T2, mmap, P, _, buffer, exec_or_read, 0).
enables_privilege_escalation(cve_2016_5195) :-
    exploited(cve_2016_5195).
\end{asplisting}

Regarding the financial fraud, we have two actions, shown in \Cref{lst:financial-fraud-pddl-actions}, one for each mechanism. For malware to perform financial fraud, it must gain access to notifications, the clipboard, or windows containing login UI fields. Thus, we have predicates for each one of those cases, some of them are inferred from the ASP as shown later in \Cref{tbl:asp-inferred-predicates} and some are enabled by performing other actions. 
\begin{figure}[!hpb]
\begin{pddllisting}[caption=Financial fraud PDDL actions, label={lst:financial-fraud-pddl-actions}]
(:action fin-fraud-mechanism-exploit
    :parameters (?a - app ?v - vuln)
    :precondition (and (exploited ?v)
       (enables-privilege-escalation ?v)
       (or (notification-accessible ?a)
        (clipboard-readable ?a)
        (login-ui-observed ?a)))
    :effect (threat-possible financial_fraud exploit ?a))

(:action fin-fraud-mechanism-permission
    :parameters (?a - app ?acc - account ?f - factor)
    :precondition (and 
    (credential-obtained ?a ?acc)
    (otp-captured ?acc ?f))
    :effect 
    (threat-possible financial_fraud permission ?a))
\end{pddllisting}
\end{figure}
Similar actions are defined for surveillance threats.
Advanced malware can perform a chain of exploitations to reach its target. To account for this we use \texttt{(pivot-exploit-from-to ?v1 - vuln ?v2 - vuln)} allow the planner to exploit a chain of vulnerabilities to achieve its goal.

\section{Evaluation \& Discussion}

We evaluate \frameworkname\footnote{Code will be publicly available upon publication.} using the KronoDroid dataset~\cite{KronoDroidDataset}, comprising 8849 real malware samples. Each contains the APK, list of API calls, permissions, metadata, system calls, and hardware information. Each sample is evaluated individually on an AMD EPYC 7763 64-Core Processor running at 2.4GHz with a time limit of 1 hour and 8GB of memory. To generate at most ten plans, we used SymK planner~\cite{speck2020symbolic}. 
\frameworkname{} was able to detect possible threats for 7331 malware samples out of 8849. Some samples (964) has timeout-ed, and \frameworkname{} did not detect any possible threats. 
The results, can be summarised as follows:
 For surveillance threats, permission-based mechanisms generated 7,204 risks resulting in 31,353 plans, while exploit-based mechanisms generated 7,114 risks resulting in 37,394 plans. For financial fraud threats, no plans were generated via permission-based mechanisms (0 risks), whereas exploit-based mechanisms generated 1,831 risks resulting in 17,810 plans.
%
%
%
%
To demonstrate \frameworkname's operation, we present a case study of the sample named ad.notify1+24240, where \frameworkname{} detected multiple potential mechanisms (e.g., permission and exploit) for surveillance. For each mechanism, \frameworkname{} generated multiple plans.
%
For the permission mechanism, \frameworkname{} inferred that several CVEs are being exploited (e.g., \texttt{cve\_2016\_5195}, \texttt{cve\_2024\_43093}) based on the system calls. This enabled the action \texttt{grant-permission-\allowbreak  for-sensors-\allowbreak mechanism-privilege-escalation}, thus allowing the planner to grant itself access to the camera sensor. 
%
%
Regarding the exploit mechanism, it detected a CVE being exploited (e.g., \texttt{cve\_2019\_2194}) which allowed the planner to exploit another CVE (e.g., \texttt{cve\_2019\_2103}) using the action \texttt{pivot-exploit}, which ultimately led to exploiting the camera sensor. 


\section{Related work}

Threat hunting has emerged as a critical proactive defence mechanism in enterprise security, and recent surveys and systematic reviews document both operational practices and research trends \cite{mahb_surv}. Industry studies report that hypothesis driven hunting and contextual analysis remain central to practitioner workflows, while academic reviews highlight progress in behaviour based detection and the limitations of purely statistical approaches. Together these works motivate richer, context aware hunting methods that fuse multiple data sources to produce more reliable and actionable hypotheses \cite{SANS}.
Research that automates hypothesis generation or ranks candidate attack explanations has advanced rapidly. \citet{nour_automa} proposed AUTOMA, an automated pipeline that generates variants of attack hypotheses from threat intelligence and telemetry using knowledge discovery techniques, explicitly producing candidate hypotheses for human analysts to validate. \citet{kaiser_attack_hypotheses} developed a method that fuses threat intelligence knowledge graphs with probabilistic reasoning to infer likely TTPs while proposing plausible attack paths from noisy evidence; their threat intelligence knowledge base demonstrates how multi level cyber threat intelligence (CTI) can be encoded and queried to produce ranked hunting hypotheses. \citet{ferdjouni2024threatscout} developed ThreatScout, an automated threat search system that leverages machine reasoning to convert telemetry and contextual signals into hunting actions and demonstrated its application in multiple profiles of threat actors.

Recent Android focused research increasingly emphasizes automatic mapping of app traces and runtime telemetry into candidate TTP for investigators. \citet{xu_dva} proposed DVa, a dynamic execution and symbolic malware analysis pipeline that extracts targeted victims, abuse vectors, and persistence mechanisms from Android accessibility malware and produces concrete, malware specific hypotheses for investigators. \citet{RARIKKAT_DroidTTP} introduced DroidTTP, which maps Android app behaviors to MITRE ATT\&CK tactics and techniques using feature engineering, machine learning, and large language models to predict TTPs from APK artifacts and runtime traces. \citet{Alam_Ladder} presented LADDER, a CTI extraction framework that derives structured attack patterns from external reports and aligns them with ATT\&CK patterns including Android relevant phases. \citet{Fairbanks_FG_android} use control flow and graph analysis to identify ATT\&CK tactics inside Android malware control flow, providing an automated path from low level code artifacts to technique level hypotheses.

\section{Conclusion}


In this paper, we introduced a threat hunting framework that automates the generation of cyber threat hypotheses using a combination of logic programming and automated planning. Since our framework is generic, we showcase it for Android devices through a set of experiments on real Android malware samples. Our framework identified possible surveillance and financial fraud threats, demonstrating that automated planners are useful as a reasoning engine for cyber threat hunting. Future work includes domain model acquisition techniques to account for novel threats based on threat intelligence reports. 
Another possible research direction is to further automate the system by exploring inductive logic learning~\cite{inductive-logic-learning} to automatically generate ASP rules.

\bibliographystyle{plainnat}  
\bibliography{ref}

\end{document}